\documentclass[twocolumn,showpacs,preprintnumbers,amsmath,amssymb,prb]{revtex4}
\usepackage{dcolumn}
\usepackage{bm}
\usepackage{amsmath}
\usepackage{amssymb}
\usepackage{graphicx}
\begin{document}
\title{Van der Waals interaction between two crossed carbon nanotubes}
\author{Alexander I. Zhbanov}
\author{Evgeny G. Pogorelov}
\email{evgeny@gate.sinica.edu.tw}
\author{Yia-Chung Chang}
\affiliation{Research Center for Applied Sciences, Academia Sinica, 128,
  Section 2, Academia Road Nankang, Taipei 115, Taiwan.}
\date{\today}

\begin{abstract}
The analytical expressions for the van der Waals potential energy and force between two crossed carbon nanotubes are presented. The Lennard-Jones potential for two carbon atoms and the method of the smeared out approximation suggested by L.A.~Girifalco were used. The exact formula is expressed in terms of rational and elliptical functions. The potential and force for carbon nanotubes were calculated. The uniform potential curves for single- and multi- wall nanotubes were plotted. The equilibrium distance, maximal attractive force, and potential energy have been estimated.
\end{abstract}
\pacs{7970+g, 8105Tp}
\maketitle

\section{INTRODUCTION}

The van der Waals (VDW) interaction plays very important role in Nano Electro Mechanical systems and Nano Electronic Devices \cite{AnantramMP2006, ZhbanovAI2004}. Carbon nanotubes (CNTs) are promising material for creation a nanotweezers \cite{KimPh1999, KeC-H2005}, nanoswitches \cite{DequesnesM2002, KinaretJM2003, RamezaniA2007}, bearings \cite{CumingsJ2000, DongL2004, DongL2006}, nanotube random access memory \cite{RueckesTh2000, KwonO-K2005}, etc. The VDW forces are very critical for understanding the growth mechanism of fullerenes and nanotubes \cite{DoyeJEK1995, SinnottSB1999, SohnJI2002} and formation process of ropes and bundles \cite{TersoffJ1994, HenrardL1999, SauvajolJ-L2002}. Potentials for graphite layers \cite{GirifalcoLA1956, AllersW1999}, two fullerenes \cite{GirifalcoLA1992, KniazK1995, GuerinH1998, BaowanD2007a}, fullerene and surface \cite{ReyC1997, GuoS2007}, nanotube and surface \cite{DongL2004, DrummondND2007, SqueSJ2007}, fullerenes inside and outside of nanotubes \cite{GirifalcoLA2000,MickelsonW2003, BaowanD2007b, CoxBJ2008, ThamwattanaN2008} are well studied. There are a number of publications devoted to interaction between the inner and the outer parallel tubes such as single- (SWNTs) \cite{GirifalcoLA2000,SunCh-H2005, CaoD2007, PopescuA2008}, double- \cite{SaitoR2001, BaowanD2007c}, and multi-wall nanotubes (MWNTs) \cite{XiaoT2004, SunCh-H2006, ZhengQ2002}.

The present work is dedicated to interplay between two CNTs crossed at arbitrary angle. Problems related with application of molecular dynamics and density functional theory for such kind of calculations are discussed for example in Refs.~\onlinecite{GirifalcoLA2000} and \onlinecite{PopescuA2008}.

To evaluate potential between two crossed SWNTs or MWNTs we apply the continuum Lenard-Jones (LJ) model suggested by L.A. Girifalco
\cite{GirifalcoLA1992}.

The model potentials for the VDW interaction are based on empirical functions whose parameters are obtained from experiment. It is remarkable that they have been so successful in providing a unified, consistent description of the properties that depend on the weak interactions between and among graphene sheets, fullerene molecules, and nanotubes.

\section{ANALYTICAL APPROACH}
\subsection{Model}

The LJ potential for two carbon atoms in graphene-graphene structure is
\begin{equation}
\varphi(r)=-\frac{A}{r^6}+\frac{B}{r^{12}}.
\end{equation}
where $r$ is a distance, $A=15.2\,\mathrm{eV}\cdot\mbox{\AA}\rule{-2pt}{3pt}^6$ and $B=24100\,\mathrm{eV}\cdot\mbox{\AA}\rule{-2pt}{3pt}^{12}$ are the attractive and repulsive constants respectively \cite{GirifalcoLA2000}.
We approximate the potential between two crossed SWNTs by integration of LJ potential
\begin{equation}
\varphi_{tt}=\nu^2\int\varphi(r)\,d\Sigma_1\,d\Sigma_2.
\end{equation}
Mean surface density of carbon atoms for hexagonal structure is
\begin{equation}
\nu=\frac{4}{3\sqrt{3}a_1^2}\approx 0.393 \mbox{ atom per \AA}^2,
\end{equation}
where $a_1=1.42\mbox{\AA}$ is observed value of the C-C bond lengths for periodic graphite\cite{SaitoR2001}. If we know van der Waals interaction between two SWNTs then we may obtain interaction between MWNTs by summation over all pairs of layers \cite{SunCh-H2006}.

\subsection{Ancillary integrals}

To calculate the VDW interaction between two nanotubes we have to take few useful integrals.
For integral of LJ potential between two straight lines
\begin{equation}
I_{ll}(r)=\int\varphi(r)\,dl_1\,dl_2,
\end{equation}
we obtain
\begin{equation}
I_{ll}(r)=\frac{\pi}{\sin\gamma}\biggl(-\frac{A}{2r^4}+
\frac{B}{5r^{10}}\biggr),
\end{equation}
where $\gamma$ is angle, $r$ is distance between lines, and index $ll$ means ``line-line''.
For next integral between line and tube we have
\begin{equation}
I_{lt}(r,t)=t\int_{-\pi}^\pi I_{ll}(r-t\sin\beta) d\beta,\;r>t,
\end{equation}
where $t$ is tube radius, $r$ is distance between line and axis of tube, and index $lt$ means
``line-tube''. Figure 1a illustrates the schematic image of SWNT and line.
\begin{figure}
\includegraphics[width=200pt]{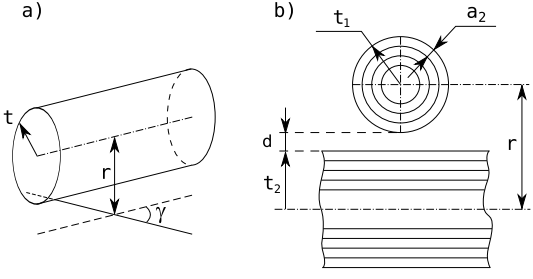}
\caption{Schematic drawing of interaction between: a) SWNT and line; b) two MWNTs}
\end{figure}
Introducing new variable $u=\tan(\beta/2)$ and using method of partial fractions we get
\begin{multline}
I_{lt}(r,t)=\frac{1}{\sin\gamma}\biggl(-\frac{A\cdot \varkappa^3 G_A(\varkappa)}{2r^3}+
\frac{B\cdot \varkappa^9 G_B(\varkappa)}{5r^9}\biggr)\\
=\frac{1}{\sin\gamma}\biggl(-\frac{A\cdot G_A(\varkappa)}{2t^3}+
\frac{B\cdot G_B(\varkappa)}{5t^9}\biggr),
\end{multline}
where
\begin{gather*}
\varkappa\equiv\frac{r}{t}\,,\;
G_A(\varkappa)=\frac{\pi^2\varkappa(2\varkappa^2+3)}{(\varkappa^2-1)^\frac72}\,,\\
G_B(\varkappa)=
\frac{\pi^2\varkappa(128\varkappa^8\!+\!2304\varkappa^6\!+\!6048\varkappa^4\!+\!3360\varkappa^2\!+\!315)}
{64(\varkappa^2-1)^\frac{19}2}\,.
\end{gather*}

\subsection{Tube-tube interaction potential}

Figure 1b schematically shows the interaction between two MWNTs crossed at right angle.
Parameter $a_2=3.44\mbox{\AA}$ is the average distance between two layers in MWNTs\cite{SaitoR2001}, $d$ is the gap between tubes.

In particular case these tubes may consist only of one layer.
We notice $t_1$ as a radius of first SWNT, and $t_2$ as a radius of second one ($r=d+t_1+t_2$). The interaction potential between two SWNTs is
\begin{equation}
\label{pot_tt}
\varphi_{tt}(r,t_1,t_2)=2\nu^2\int_{r-t_2}^{r+t_2} I_{lt}(x,t_1)\sqrt{1+y'^2(x)}\, dx,
\end{equation}
where $y(x)=\sqrt{t_2^2-(x-r)^2}$, index $tt$ means ``tube-tube''.

Introducing dimensionless parameters
\begin{equation}
\varkappa=\frac{x}{t_1},\;b_1=\frac{r}{t_1},\;b_2=\frac{r}{t_2},\;k=\frac{t_2}{t_1}\,,
\end{equation}
and using (7) we have
\begin{equation}
\varphi_{tt}(r,t_1,t_2)=\nu^2 t_1\int_{b_1-k}^{b_1+k}
\frac{2kI_{lt}(\varkappa,t_1)}{\sqrt{k^2-(\varkappa-b_1)^2}}\,d\varkappa.
\end{equation}
After some transformations we write
\begin{equation}
\label{pot_tt}
\varphi_{tt}(r,t_1,t_2)=\frac{\nu^2}{\sin\gamma}
\biggl(-\frac{A\cdot g_A}{r^2}+\frac{B\cdot g_B}{r^8}\biggr),
\end{equation}
where multipliers for attractive and repulsive terms are
\begin{gather}
g_A(b_1,b_2)=\frac{b_1^2}{2}\int_{b_1-k}^{b_1+k} \frac{2kG_A(\varkappa)}{\sqrt{k^2-(\varkappa-b_1)^2}}\, d\varkappa,\\
g_B(b_1,b_2)=\frac{b_1^8}{5}\int_{b_1-k}^{b_1+k} \frac{2kG_B(\varkappa)}{\sqrt{k^2-(\varkappa-b_1)^2}}\, d\varkappa.
\end{gather}

The obtained Eqs.~(12,~13) represents the elliptic integrals.
In modern mathematics elliptic integral is defined as integral $\int R(x,y)dx$, where $R(x,y)$ is rational function of $x$ and $y$,
and $y^2$ is a qubic or quartic polynomial in $x$. With the apropriate reduction formula every elliptical integral can be expressed in terms of elementary functions and canonical elliptic integrals of first, second and third kind. The method of integration is quite complicated but well known \cite{Abramowitz1964, Hancock1910, Greenhill1892, King1924}.
We would like to present only final answer. For attractive part we have the dimensionless parameter
\begin{equation}
g_A=g_{AK}K(h)+g_{AE}E(h),
\end{equation}
where
\begin{equation}
h=\frac{2\sqrt{b_1b_2}}{\sqrt{(b_1b_2+b_1-b_2)(b_1b_2+b_2-b_1)}},
\end{equation}
\begin{multline}
g_{AK}=-\Bigl[2\pi^2b_1^4b_2^4\sum_{i,j=1..3}\{p_{AK}\}_{ij}b_1^{2(i-1)}b_2^{2(j-1)}
\Bigr]\\
/\Bigl[3(b_1b_2+b_1+b_2)^2(b_1b_2-b_1-b_2)^2(b_1b_2+b_1-b_2)^{\frac52}\\
\times (b_1b_2+b_2-b_1)^{\frac52}\Bigr],
\end{multline}
\begin{multline}
g_{AE}=\Bigl[2\pi^2b_1^4b_2^4\sum_{i,j=1..4}\{p_{AE}\}_{ij}b_1^{2(i-1)}b_2^{2(j-1)}
\Bigr]\\
/\Bigl[3(b_1b_2+b_1+b_2)^3(b_1b_2-b_1-b_2)^3(b_1b_2+b_1-b_2)^{\frac52}\\
\times (b_1b_2+b_2-b_1)^{\frac52}\Bigr],
\end{multline}
and matrixes of integer coefficients are
\begin{equation*}
\{p_{AK}\}=
\begin{bmatrix}
0&0&-3\\0&6&-2\\-3&-2&5
\end{bmatrix},
\end{equation*}
\begin{equation*}
\{p_{AE}\}=
\begin{bmatrix}
0&0&0&12\\0&0&-12&-13\\0&-12&58&-10\\12&-13&-10&11
\end{bmatrix}.
\end{equation*}
Analogically for repulsive part we write the dimensionless parameter
\begin{equation}
g_B=g_{BK}K(h)+g_{BE}E(h),
\end{equation}
where
\begin{multline}
g_{BK}=-\Bigl[
\pi^2b_1^{10}b_2^{10}\!\!\sum_{i,j=1..12}\!\!
\{p_{BK}\}_{ij}b_1^{2(i-1)}b_2^{2(j-1)}
\Bigr]\\
/\Bigl[6300(b_1b_2+b_1+b_2)^8(b_1b_2-b_1-b_2)^8\\
\times (b_1b_2+b_1-b_2)^{\frac{17}2}(b_1b_2+b_2-b_1)^{\frac{17}2}\Bigr],
\end{multline}
\begin{multline}
g_{BE}=\Bigl[
\pi^2b_1^{10}b_2^{10}\sum_{i,j=1..13}\{p_{BE}\}_{ij}b_1^{2(i-1)}b_2^{2(j-1)}
\Bigr]\\
/\Bigl[25200(b_1b_2+b_1+b_2)^9(b_1b_2-b_1-b_2)^9\\
\times (b_1b_2+b_1-b_2)^{\frac{17}2}(b_1b_2+b_2-b_1)^{\frac{17}2}\Bigr].
\end{multline}

\begin{table*}
\caption{Matrix $\{p_{BK}\}$}
\begin{equation*}
\left[
{\small
\begin{array}{cccccccccccc}
0&0&0&0&0&0&0&0&0&0&0&-19530\\
\phantom{0}&0&0&0&0&0&0&0&0&0&14490&-104685\\
\phantom{0}&\phantom{0}&0&0&0&0&0&0&0&445410&-2015790&765744\\
\phantom{0}&\phantom{0}&\phantom{0}&0&0&0&0&0&-1760850&7748055&-2248848&-1029636\\
\phantom{0}&\phantom{0}&\phantom{0}&\phantom{0}&0&0&0&2758140&
-938280&-36368640&27011520&-1356976\\
\phantom{0}&\phantom{0}&\phantom{0}&\phantom{0}&\phantom{0}&0&
-1437660&-31109610&114941568&-34271280&-35205136&5012854\\
\phantom{0}&\phantom{0}&\phantom{0}&\phantom{0}&\phantom{0}&\phantom{0}&
52840620&-77089824&-172119744&193452624&-11269956&-5001276\\
\phantom{0}&\phantom{0}&\phantom{0}&\phantom{0}&\phantom{0}&\phantom{0}&\phantom{0}&
360818280&-161478032&-132247830&47389620&1327256\\
\phantom{0}&\phantom{0}&\phantom{0}&\phantom{0}&\phantom{0}&\phantom{0}&\phantom{0}&
\phantom{0}&
295052744&-50123640&-24636352&1062586\\
\phantom{0}&\phantom{0}&\phantom{0}&\phantom{0}&\phantom{0}&\phantom{0}
\lefteqn{\mbox{Symmetry}}
&\phantom{0}&
\phantom{0}&\phantom{0}&
54162000&-2009850&-786849\\
\phantom{0}&\phantom{0}&\phantom{0}&\phantom{0}&\phantom{0}&\phantom{0}&\phantom{0}&
\phantom{0}&\phantom{0}&\phantom{0}&
2858226&112076\\
\phantom{0}&\phantom{0}&\phantom{0}&\phantom{0}&\phantom{0}&\phantom{0}&\phantom{0}&
\phantom{0}&\phantom{0}&\phantom{0}&\phantom{0}&
18436
\end{array}
}
\right]
\end{equation*}
\end{table*}

\begin{table*}
\caption{Matrix $\{p_{BE}\}$}
\begin{equation*}
{\footnotesize
\left[\;
\begin{array}{ccccccccccccc}
0\!&\!0\!&\!0\!&\!0\!&\!0\!&\!0\!&\!0\!&\!0\!&\!0\!&\!0\!&\!0\!&\!0\!&\!177345\\
\phantom{0}\!&\!0\!&\!0\!&\!0\!&\!0\!&\!0\!&\!0\!&\!0\!&\!0\!&\!0\!&\!0\!&\!
976500\!&\!605220\\
\phantom{0}\!&\!\phantom{0}\!&\!0\!&\!0\!&\!0\!&\!0\!&\!0\!&\!0\!&\!0\!&\!0\!&\!
-9019710\!&\!28878780\!&\!-7265496\\
\phantom{0}\!&\!\phantom{0}\!&\!\phantom{0}\!&\!0\!&\!0\!&\!0\!&\!0\!&\!0\!&\!
0\!&\!18117540\!&\!-26421780\!&\!-43621536\!&\!16744920\\
\phantom{0}\!&\!\phantom{0}\!&\!\phantom{0}\!&\!\phantom{0}\!&\!0\!&\!0\!&\!
0\!&\!0\!&\!4242735\!&\!-322255500\!&\!735299208\!&\!-256914792\!&\!-3437690\\
\phantom{0}\!&\!\phantom{0}\!&\!\phantom{0}\!&\!\phantom{0}\!&\!\phantom{0}\!&\!
0\!&\!0\!&\!-66510360\!&\!827431080\!&\!-863979648\!&\!-1027666080\!&\!
746908112\!&\!-43635920\\
\phantom{0}\!&\!\phantom{0}\!&\!\phantom{0}\!&\!\phantom{0}\!&\!\phantom{0}\!&\!
\phantom{0}\!&\!104031900\!&\!-508237800\!&\!-2282192304\!&\!5778243552\!&\!
-1895641688\!&\!-527458544\!&\!77409220\\
\phantom{0}\!&\!\phantom{0}\!&\!\phantom{0}\!&\!\phantom{0}\!&\!\phantom{0}\!&\!
\phantom{0}\!&\!\phantom{0}\!&\!4923519552\!&\!-4510407600\!&\!-4710377296\!&\!
4534092720\!&\!-365372760\!&\!-54614168\\
\phantom{0}\!&\!\phantom{0}\!&\!\phantom{0}\!&\!\phantom{0}\!&\!\phantom{0}\!&\!
\phantom{0}\!&\!\phantom{0}\!&\!\phantom{0}\!&\!12018698404\!&\!-4283796976\!&\!
-2131197060\!&\!690518216\!&\!8213905\\
\phantom{0}\!&\!\phantom{0}\!&\!\phantom{0}\!&\!\phantom{0}\!&\!\phantom{0}\!&\!
\phantom{0}\!&\!\phantom{0}\!&\!\phantom{0}\!&\!\phantom{0}\!&\!5458820144\!&\!
-809953328\!&\!-276097028\!&\!11278540\\
\phantom{0}\!&\!\phantom{0}\!&\!\phantom{0}\!&\!\phantom{0}\!&\!\phantom{0}\!&\!
\phantom{0}\!&\!\phantom{0}\lefteqn{\mbox{Symmetry}}
\!&\!\phantom{0}\!&\!\phantom{0}\!&\!\phantom{0}\!&\!
660421350\!&\!-23545932\!&\!-6367700\\
\phantom{0}\!&\!\phantom{0}\!&\!\phantom{0}\!&\!\phantom{0}\!&\!\phantom{0}\!&\!
\phantom{0}\!&\!\phantom{0}\!&\!\phantom{0}\!&\!\phantom{0}\!&\!\phantom{0}\!&\!
\phantom{0}\!&\!24951224\!&\!777760\\
\phantom{0}\!&\!\phantom{0}\!&\!\phantom{0}\!&\!\phantom{0}\!&\!\phantom{0}\!&\!
\phantom{0}\!&\!\phantom{0}\!&\!\phantom{0}\!&\!\phantom{0}\!&\!\phantom{0}\!&\!
\phantom{0}\!&\!\phantom{0}\!&\!114064
\end{array}
\right]
}
\end{equation*}
\end{table*}

Matrixes of integer coefficients $\{p_{BK}\}$ and $\{p_{BE}\}$ are placed in Table I and II respectively.

As we see the final result (\ref{pot_tt}) is quite huge but it is working much better then usual numerical integration, because this analytical formula provides high accuracy and high speed of calculations.

\subsection{Tube-tube force}

The VDW resulting force caused by VDW interaction energy is
\begin{equation}
F(r)=-\frac{d\varphi_{tt}(r)}{dr}.
\end{equation}

Using expressions\cite{Hancock1910}
\begin{equation}
\frac{d K(x)}{dx}=\frac{E(x)}{(1-x^2)x}-\frac{K(x)}{x}
\end{equation}
and
\begin{equation}
\frac{dE(x)}{dx}=\frac{1}{x}\bigl(E(x)-K(x)\bigr),
\end{equation}
it is possible to obtain the analytical formula for VDW force. After usual differentiation over $r$ we have
\begin{equation}
F(r,t_1,t_2)=\frac{\nu^2}{\sin\gamma}\biggl(
-\frac{A\cdot f_A}{r^3}+\frac{B\cdot f_B}{r^9}\biggr),
\end{equation}
where
\begin{gather}
f_A=f_{AK}K(h)+f_{AE}E(h),\\
f_B=f_{BK}K(h)+f_{BE}E(h).
\end{gather}
The dimensionless coefficients for attractive and repulsive part of force are expressed as
\begin{multline}
f_{AK}=\Bigl[2\pi^2b_1^4b_2^4\sum_{i,j=1..5}\{q_{AK}\}_{ij}b_1^{2(i-1)}b_2^{2(j-1)}
\Bigr]\\
/\Bigl[3(b_1b_2+b_1+b_2)^3(b_1b_2-b_1-b_2)^3(b_1b_2+b_1-b_2)^{\frac72}\\
\times (b_1b_2+b_2-b_1)^{\frac72}\Bigr],
\end{multline}
\begin{multline}
f_{AE}=-\Bigl[4\pi^2b_1^4b_2^4\sum_{i,j=1..6}\{q_{AE}\}_{ij}b_1^{2(i-1)}b_2^{2(j-1)}
\Bigr]\\
/\Bigl[3(b_1b_2+b_1+b_2)^3(b_1b_2-b_1-b_2)^3(b_1b_2+b_1-b_2)^{\frac72}\\
\times (b_1b_2+b_2-b_1)^{\frac72}(b_1^2b_2^2-2b_1b_2-b_1^2-b_2^2)\Bigr],
\end{multline}
\begin{multline}
f_{BK}=\Bigl[\pi^2b_1^{10}b_2^{10}\sum_{i,j=1..14}\{q_{BK}\}_{ij}
b_1^{2(i-1)}b_2^{2(j-1)}
\Bigr]\\
/\Bigl[5040(b_1b_2+b_1+b_2)^9(b_1b_2-b_1-b_2)^9(b_1b_2+b_1-b_2)^{\frac{19}2}\\
\times (b_1b_2+b_2-b_1)^{\frac{19}2}\Bigr],
\end{multline}
\begin{multline}
f_{BE}=-\Bigl[\pi^2b_1^{10}b_2^{10}\sum_{i,j=1..15}\{q_{BE}\}_{ij}b_1^{2(i-1)}b_2^{2(j-1)}
\Bigr]\\
/\Bigl[5040(b_1b_2+b_1+b_2)^9(b_1b_2-b_1-b_2)^9(b_1b_2+b_1-b_2)^{\frac{19}2}\\
\times (b_1b_2+b_2-b_1)^{\frac{19}2}(b_1^2b_2^2-2b_1b_2-b_1^2-b_2^2)\Bigr],
\end{multline}
where
\begin{equation*}
\{q_{AK}\}=
\begin{bmatrix}
0&0&0&0&3\\
0&0&0&-12&36\\
0&0&18&-36&-55\\
0&-12&-36&158&-10\\
3&36&-55&-10&26
\end{bmatrix},
\end{equation*}
\begin{equation*}
\{q_{AE}\}=
\begin{bmatrix}
0&0&0&0&0&-6\\
0&0&0&0&18&-39\\
0&0&0&-12&-132&128\\
0&0&-12&342&-224&-90\\
0&18&-132&-224&356&-18\\
-6&-39&128&-90&-18&25
\end{bmatrix}.
\end{equation*}
Matrixes $\{q_{BK}\}$ and $\{q_{BE}\}$ are given in Table III and IV respectively.
\begin{table*}
\caption{Matrix $\{q_{BK}\}$}
\begin{equation*}
\left[\;
\begin{smallmatrix}
0& 0& 0& 0& 0& 0& 0& 0& 0& 0& 0& 0& 0& 15624\\
& 0& 0& 0& 0& 0& 0& 0& 0& 0& 0& 0& -42840& 755433\\
&& 0& 0& 0& 0& 0& 0& 0& 0& 0& -317520& 4651668& -1248828\\
&&& 0& 0& 0& 0& 0& 0& 0& 2109744& -40861422& 65362332& -11592924\\
&&&& 0& 0& 0& 0& 0& -5380200& 79474500& 10951500& -192184440& 44486040\\
&&&&& 0& 0& 0& 6971832& 28477575& -991483500& 1506696660& -240160680& -50195242\\
&&&&&& 0& -3356640& -318885336& 2273850600& -1148679840& -2776597920& 1286509520& -21785104\\
&&&&&&& 492775164& -1357432104& -5600776440& 10812253920& -1915609240& -1300087984& 107776844\\
&&&&&&&& 10893073968& -7839981360& -9567436112& 7173492336& -138383496& -103517008\\
&&&&&&&&& 20757703300& -6144532400& -3892023180& 951956720& 33353705\\
&&&&&&&&&& 8420997520& -1012896160& -459154660& 9346988\\
&&&&&&&\lefteqn{\mbox{Symmetry}}&&&& 973066230& -17001900& -8899384\\
&&&&&&&&&&&& 37187456& 1348304\\
&&&&&&&&&&&&& 155552
\end{smallmatrix}
\right]
\end{equation*}
\end{table*}

\begin{table*}
\caption{Matrix $\{q_{BE}\}$}
\begin{equation*}
\left[\;
\begin{smallmatrix}
0& 0& 0& 0& 0& 0& 0& 0& 0& 0& 0& 0& 0& 0& -35469\\
& 0& 0& 0& 0& 0& 0& 0& 0& 0& 0& 0& 0& -124362& -1553454\\
&& 0& 0& 0& 0& 0& 0& 0& 0& 0& 0& 2159073& -24489990& 4928091\\
&&& 0& 0& 0& 0& 0& 0& 0& 0& -7426692& 104520780& -139540884& 17567676\\
&&&& 0& 0& 0& 0& 0& 0& 8202411& 84564396& -901760874& 861398916&
-110403234\\
&&&&& 0& 0& 0& 0& 11375658& -1024742250& 4609741500& -3041479980&
-594533340& 204320812\\
&&&&&& 0& 0& -48259071& 1921604958& -3163636875& -12201112980&
16321695750& -3128825524& -103004658\\
&&&&&&& 68216904& -1059904440& -11093841192& 45974714520&
-25761560400& -14217370384& 6921793360& -196048368\\
&&&&&&&& 21368220468& -31611088152& -55358109540& 87874942896& -17650761720& -4488366288& 380618007\\
&&&&&&&&& 131005821528& -71790032408& -50711821008& 34135947792& -1350545098& -270248958\\
&&&&&&&&&& 128236620500& -32263986000& -13793389735& 3110622970& 63736407\\
&&&&&&&&&&& 33152425140& -3602984700& -1197711564& 24929412\\
&&&&&&&&\lefteqn{\mbox{Symmetry}}&&&& 2709914346& -49166500& -17323848\\
&&&&&&&&&&&&& 77206912& 2281392\\
&&&&&&&&&&&&&& 236192
\end{smallmatrix}
\right]
\end{equation*}
\end{table*}

\subsection{Potential and force between equivalent tubes}

\begin{figure}
\includegraphics[width=200pt]{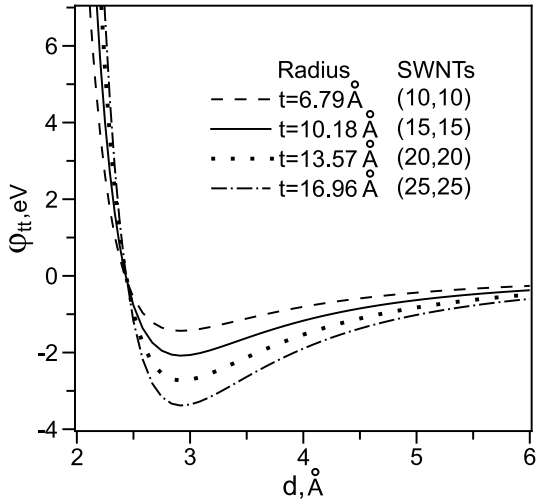}
\caption{Potential energies for interaction between pairs of identical SWNTs}
\end{figure}

\begin{figure}
\includegraphics[width=200pt]{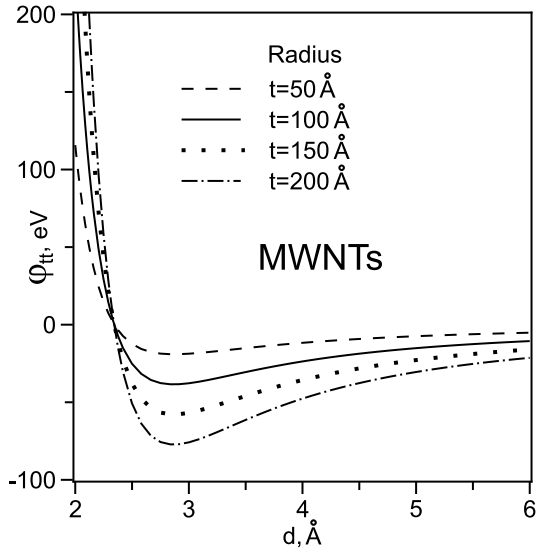}
\caption{Potential energies for interaction between pairs of MWNTs of equivalent size.
MWNTs contain exactly 10 layers}
\end{figure}

\begin{figure}
\includegraphics[width=180pt]{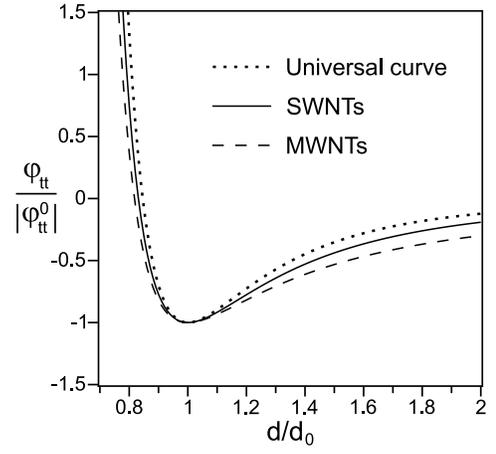}
\caption{Uniform potential for CNTs with arbitrary sizes. Dotted line is the universal curve suggested by L.A.~Girifalco et al.}
\end{figure}

\begin{figure}
\includegraphics[width=200pt]{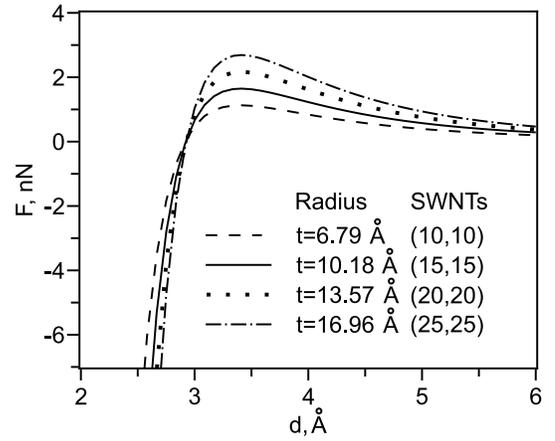}
\caption{van der Waals forces between two identical SWNTs}
\end{figure}

\begin{figure}
\includegraphics[width=200pt]{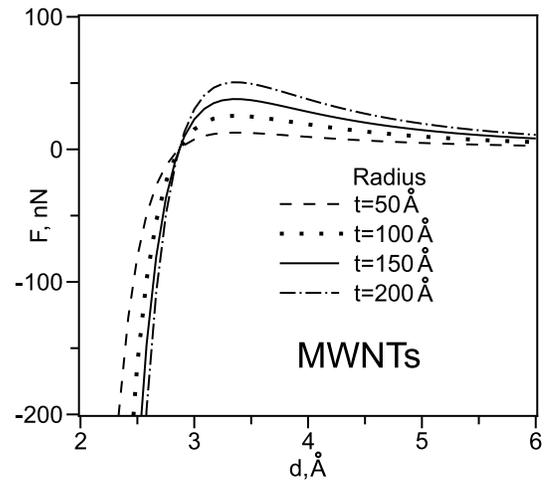}
\caption{van der Waals force between pairs of MWNTs of equivalent size.
MWNTs contain 10 layers}
\end{figure}

\begin{table}
\caption{Dependence of potential well $|\varphi_{tt}^0|$ (eV) from number of layers for MWNTs of equivalent radii}
\begin{tabular}{lccccc}
\hline\hline\\
Radius (\AA) & 5 & 10 & 15 & 20 & 25\\
\hline\\
50 & 18.57 & 19.05 & N/A & N/A & N/A \\
100 & 37.31 & 38.46 & 38.84 & 39.00 & 39.08\\
150 & 56.06 & 57.85 & 58.48 & 58.78 & 58.95 \\
200 & 74.81 & 77.25 & 78.11 & 78.54 & 78.79 \\
\hline\hline
\end{tabular}
\end{table}

\begin{table*}
\caption{Calculated depth $|\varphi_{tt}^0|$ (eV) for SWNTs (upper-right side) and approximation (lower-left side)}
\begin{tabular}{ccccc}
\hline\hline\\
Tube type & (10,10) & (15,15) & (20,20) & (25,25) \\
Radius (\AA) & 6.79 & 10.18 & 13.57 & 16.96\\
\hline\\
(10,10) 6.79 & 1.426 $\backslash$ 1.434 & 1.728 & 1.979 & 2.202\\
(15,15) 10.18 & 1.723 & 2.080 $\backslash$ 2.081 & 2.384 & 2.653\\
(20,20) 13.57 & 1.975 & 2.385 & 2.734 $\backslash$ 2.731 & 3.039\\
(25,25) 16.96 & 2.199 & 2.655 & 3.043 & 3.388 $\backslash$ 3.382\\
\hline\hline
\end{tabular}
\end{table*}

\begin{table*}
\caption{Calculated depth $|\varphi_{tt}^0|$ (eV) for MWNTs consisting of 10 shells (upper-right side) and approximation (lower-left side)}
\begin{tabular}{ccccc}
\hline\hline\\
Radius (\AA) & 50 & 100 & 150 & 200\\
\hline\\
50 & 19.03 $\backslash$ 19.05 & 27.07 & 33.20 & 38.36\\
100 & 27.04 & 38.43 $\backslash$ 38.46 & 47.17 & 54.51\\
150 & 33.17 & 47.14 & 57.83 $\backslash$ 57.85 & 66.85\\
200 & 38.33 & 54.48 & 66.83 & 77.23 $\backslash$ 77.25\\
\hline\hline
\end{tabular}
\end{table*}

The potential and force can be expressed essentially simpler in the case when radii of interacted tubes are equal $t_1=t_2=t$, therefore $b_1=b_2=b=r/t$.
Then we have for potential
\begin{equation}
\varphi^*_{tt}(r,t)=\frac{\nu^2}{\sin\gamma}
\biggl(-\frac{A\cdot g^*_A}{r^2}+\frac{B\cdot g^*_B}{r^8}\biggr),
\end{equation}
where
\begin{gather}
g^*_A=g^*_{AK}K(2/b)+g^*_{AE}E(2/b),\\
g^*_B=g^*_{BK}K(2/b)+g^*_{BE}E(2/b),
\end{gather}
and
\begin{equation}
g^*_{AK}=-\frac{2\pi^2(5b^2-4)}{3(b^2-4)^2}\,,
\end{equation}
\begin{equation}
g^*_{AE}=\frac{2\pi^2(32-20b^2+11b^4)}{3(b^2-4)^3}\,,
\end{equation}
\begin{multline}
g^*_{BK}=-\pi^2(4609b^{14}+56038b^{12}+321132b^{10}\\
-473632b^8+1885952b^6-3867648b^4+4510720b^2\\
-2293760)/(1575(b^2-4)^8),
\end{multline}
\begin{multline}
g^*_{BE}=\pi^2(7129b^{16}+97220b^{14}+763489b^{12}\\
-1533424b^{10}+7790944b^8-21756160b^6+38781184b^4\\
-40099840b^2+18350080)/(1575(b^2-4)^9).
\end{multline}
By totally the same way in the case of $t_1=t_2$ for force we have
\begin{equation}
F^*(t,r)=\frac{\nu^2}{\sin\gamma}\biggl(
-\frac{Af^*_A}{t^2}+\frac{Bf^*_B}{t^8}\biggr),
\end{equation}
\begin{equation}
f^*_A=\frac{4\pi^2}{3b^3(b^2-4)^4}\biggl(
f^*_{AK}K\biggl(\frac{2}{b}\biggr)+f^*_{AE}E\biggl(\frac{2}{b}\biggr)
\biggr),
\end{equation}
\begin{equation}
f^*_B=\frac{\pi^2}{315b^9(b^2-4)^{10}}\biggl(
f^*_{BK}K\biggl(\frac{2}{b}\biggr)+f^*_{BE}E\biggl(\frac{2}{b}\biggr)
\biggr),
\end{equation}
\begin{equation}
f^*_{AK}=13b^6-62b^4+64b^2-96,
\end{equation}
\begin{equation}
f^*_{AE}=-25b^6+36b^4-176b^2+192,
\end{equation}
\begin{multline}
f^*_{BK}=9722b^{18}+129650b^{16}+537641b^{14}\\
-5804036b^{12}+11418976b^{10}-50923136b^8\\
+118211840b^6-180548608b^4\\
+162971648b^2-66060288,
\end{multline}
\begin{multline}
f^*_{BE}=-14762b^{18}-285174b^{16}-2659951b^{14}\\
+3029636b^{12}-27622752b^{10}+95326336b^8\\
-226611968b^6+351556608b^4\\
-321814528b^2+132120576.
\end{multline}

\section{RESULTS AND DISCUSSION}
We have studied the WDV interaction between two crossed CNTs by using the continuum LJ approximation. The analytical integrations for the potential energy of interaction between two identical SWNTs are plotted in Fig. 2. We assume nanotubes are crossed at a right angle in all our following illustrations and calculations both for SWNTs and MWNTs. We use parameter $d=r-2t$ to characterize the distance between tubes. Based on the results illustrated in Fig. 2, it can be concluded that real gap between surfaces of interacting SWNTs in equilibrium state is changed slightly in range $d=2.92..2.93$ \AA. This distance is practically independent from the angle of nanotube intersection and the diameter proportion. For comparison in Ref. \onlinecite{GirifalcoLA1992} the equilibrium gap between two fullerenes ${\rm C}_{60}$ is given as 2.95 \AA.

In the case of MWNTs interaction we apply another attractive ($A=18.6 \mbox{eV}\cdot\mbox{\AA}^6$) and repulsive ($B=29040 \mbox{eV}\cdot\mbox{\AA}^{12}$) constants which reproduce the layer distance of 3.35 {\AA} and the elastic constant $C_{33}=4.08\mbox{ GPa}$ of graphite \cite{SaitoR2001}. We assume that each pair of layers interacts as SWNTs and use summation over all pairs. The potential energy for two MWNTs of equivalent radii is plotted in Fig. 3. In these calculations we assume that each MWNT consists exactly of 10 walls. The equilibrium distance between their surfaces is found to be $d_0=2.87\mbox{ \AA}$, which is smaller then the equilibrium SWNT-SWNT gap. From our calculations follows that only several outer shells of MWNTs play essential role in the VDW interaction. For example, if two equal MWNTs with $d=200\mbox{ \AA}$ contain 5, 10, 15, 20 or 25 layers then the minimum energy is -74.8, -77.2, -78.1, -78.5 or -78.8 eV respectively.
Dependence of the minimum potential energy from number of inner layers for pairs of equivalent MWNTs is presented in Table V.

It was found that the VDW interaction between ${\rm C}_{60}-{\rm C}_{60}$, ${\rm C}_{60}$-SWNT, ${\rm C}_{60}$-graphene, graphene-graphene, parallel SWNT-SWNT, and parallel MWNT-MWNT can be described by a universal curve \cite{GirifalcoLA2000, SunCh-H2005, SunCh-H2006}. In our case universal curve means that a plot of $\overline\varphi_{tt}=\varphi_{tt}/|\varphi_{tt}^0|$ against $\overline d=d/d_0$ gives the same curve for all tube-tube interactions, where $\varphi_{tt}^0$ is the minimum energy and $d_0$ is the equilibrium spacing for the two crossed tubes. As pointed above, the equilibrium distances are approximately constants both for SWNTs and MWNTs.

We have calculated the minimum energy $\varphi^0_{tt}$ for SWNTs of different radii (Table VI) as well as for MWNTs (Table VII). These results can be described by approximating formula
\begin{equation*}
\varphi^0_{tt}=C^{CNT}_1\sqrt{t_1t_2}+C^{CNT}_2\,\frac{t_1+t_2}{\sqrt{t_1t_2}},
\end{equation*}
where $C^{SWNT}_1=-0.19285\mbox{eV}\cdot\mbox{\AA}^{-1}$, $C^{SWNT}_2=-0.05847\mbox{eV}$, and $C^{MWNT}_1=-0.388\mbox{eV}\cdot\mbox{\AA}^{-1}$, $C^{MWNT}_2=0.186\mbox{eV}$ are parameters for SWNTs and MWNTs respectively. It can be figured out from Tables VI and VII that approximating formula gives very good accuracy.
Using dimensionless potential $\overline\varphi_{tt}$ we can fit the potential of interaction between pairs of different SWNTs to one uniform curve and between pairs of different MWNTs to another one (Fig. 4).

It is remarkable that plots for CNTs of the different radii fall down in the corresponding curves with accuracy of line thickness. For comparison Fig.~4 also shows a universal potential suggested by Girifalco L.A. et al. \cite{GirifalcoLA2000}.

Figs. 5 and 6 show forces for two CNTs of equivalent radii. As we see in figures the behavior both for SWNTs and for MWNTs is qualitatively similar. The distance where attractive force reaches its maximum is in range 3.40-3.41 for SWNTs and it is practically constant, 3.36 {\AA} for MWNTs.

\section{SUMMARY AND CONCLUSIONS}

In summary, we used Lennard-Jones potential for two carbon atoms and apply method of the
smeared out approximation suggested by L.A. Girifalco to calculate interaction between two
crossed CNTs of uniform and different diameters. The exact formulas for potential energy
and van der Waals forces are expressed in terms of rational and elliptical functions.
These formulas become essentially simpler in the case of interaction between equivalent tubes.
We estimated the equilibrium distance, maximal attractive force and potential energy for SWNTs and MWNTs. We plotted uniform potential for SWNTs and MWNTs.

\section{ACKNOLEGMENTS}
   We gratefully acknowledge support through the National Science Council of Taiwan,
Republic of China, through the project NSC 95-2112-M-001-068-MY3.

\bibliography{nanotubes}
\end{document}